\def\be{\begin{equation}}
\def\ee{\end{equation}}
\def\ba{\begin{eqnarray}}
\def\ea{\end{eqnarray}}
\def\l{\label}
\def\n{\nonumber \\}

\documentstyle[12pt]{article}
 \begin{document} 


\title{Charge fluctuations in a quark-antiquark system}

\author{A.Bialas \\ M.Smoluchowski Institute of Physics \\Jagellonian
University, Cracow\thanks{Address: Reymonta 4, 30-059 Krakow, Poland;
e-mail:bialas@th.if.uj.edu.pl;}}
\maketitle

\begin{abstract} 
It is pointed out that the recent data on charge fluctuations observed
in heavy ion collisions are compatible with production of a system
of weakly correlated constituent quarks and antiquarks. 
\end{abstract}
It was recently suggested  \cite{mu1,jk} (see also \cite{kbj,jp,mu2})
 that measurements of the quantity 
\be
D= 4 \frac{<\delta Q^2>}{<N_{ch}>}  \l{1}
\ee
can be used to distinquish the hadron gas in equilibrium from the
quark-gluon plasma. 

It was then argued in \cite{kbj} that -after appropriate
corrections for resonance production are taken into account- $D\approx 3$
for the hadron gas whereas for the quark-gluon plasma $D\approx 1$.

The preliminary data from CERES \cite{ce}, NA49 \cite{49} and STAR
\cite{st} experiments indicate that the measured value of $D$ is close
to that predicted for hadron gas and differs markedly from that expected
for QGP, i.e. for a weakly correlated quark-gluon system. 

This rises a question: what kind of structure of the partonic system
produced at the early stage of  collisions may be compatible with
these experimental results? The present note is an attempt to answer
this question.

The first observation is that the estimate of the numerator in (\ref{1})
is rather straigthforward.  The argument
(essentially the repetition of the argument given in \cite{jk})
goes as follows. Let us consider a system of several particle species
(labelled by $i$) with charges $q_i$ and multiplicities $n_i$.
Since 
\ba
Q=\sum_i q_i n_i\;\;\;\rightarrow \;\;\;<Q>=\sum_i q_i <n_i>     \l{2}
\ea
 we obtain
\ba
<\delta Q^2>\equiv <Q^2>-<Q>^2=  \sum_i (q_i)^2 <n_i> 
+\sum_{i,k} c^{(2)}_{ik} <n_i><n_k> q_iq_k       \l{3}
\ea
where
$c^{(2)}_{ik}$ are the normalized two-particle correlation functions:
\ba
c^{(2)}_{ii}= \frac{<n_i(n_i-1)>}{<n_i>^2} -1;\;\;\;\;
c^{(2)}_{ik}= \frac {<n_in_k>}{<n_i><n_k>}-1\;\; if \;\; i\neq k\n \l{4}
\ea

If particles are uncorrelated the second term in (\ref{3}) vanishes.
Moreover, one notes that the first term is the sum of only positive
quantities, whereas in the second term cancellations are possible. For
example, if all normalized two-particle correlations are identical, one
obtains
\ba
<\delta Q^2>=  \sum_i (q_i)^2 <n_i>   +c^{(2)} <Q>^2  . \l{5}
\ea
and thus the second term disappers  when $<Q> =0$.

From now on  we shall thus restrict the discussion
 to the case when the second term can be neglected and thus
\ba
<\delta Q^2>= \sum_i (q_i)^2 <n_i> \l{6}
\ea
For pion gas this means
\ba
<\delta Q^2>=  <n_+>+<n_->= <N_{ch}>   \l{9}
\ea
and thus
\ba
D =4.           \l{12}
\ea
For a quark-gluon system we obtain
\ba
<\delta Q^2>=\frac19(4 <n_u> + <n_d> +4<n_{\bar{u}}>+<n_{\bar{d}}>)\l{10}
\ea
(gluons of course do not contribute). In the simplest case when
 abundances of all quarks are identical we have 
\ba
<\delta Q^2>= \frac5{18} <N_q>     \l{11}
\ea
where $<N_q>$ is the total number of quarks and antiquarks.

To calculate $D$ it is, however, also necessary to estimate $<N_{ch}>$
and this is of course model-dependent. The question is the role of
gluons. If they decay into $q-\bar q$ pairs before making hadrons, the
number of hadrons may be very large. This is what happens in the QGP
scenario. And this is the reason of the small value of $D$ obtained in
\cite{mu1,jk}. The argument is based on entropy. Qualitatively: since
gluons have a large entropy, they have to produce many particles. Some
of them will be charged and thus contribute to the denominator of
(\ref{1}).

If one considers the $q-\bar{q}$ coalescence scenario \cite{al, bra},
however, the result is very different. In this scenario gluons are
attached to quarks (forming constituent quarks) and thus do not provide
important contribution to entropy. Hadrons are created simply by
coalescence of the (constituent) $q-\bar{ q}$ pairs. Consequently, the
total number of hadrons is about $1/2$ of the total number of quarks and
antiquarks:
\ba
<N_h> =\frac12 <N_q>   \l{13}
\ea
It is natural to assume that $2/3$ of them are charged, so that we obtain
\ba
<N_{ch}> = \frac 23 <N_h>       = \frac13 <N_q>  . \l{14}
\ea
and thus 
\ba
D= \frac {10}3   = 3.333  \l{15}
\ea
which is only slightly smaller than (\ref{12}).

The values of $D$ given by (\ref{12}) and (\ref{15}) should still be
corrected for resonance production \cite{jk}. Since the difference
between them is fairly small, however, one should not expect a great
difference between the corrected results, either.

We thus conclude that the data on charge fluctuations are not
incompatible with the idea that a (quasi)uncorrelated constituent quark
system is created before hadronization. This observation strenghtens
earlier arguments leading to the coalescence picture of hadronization
(see e.g. \cite{al} -\cite{cs}).

Two comments are in order.

(i) The present data are given for fairly small rapidity intervals and
thus important effects of charge transport between the boundaries of the
considered phase-space region may be present \cite{br,je}, destroying
the argument of Refs. \cite{mu1} and \cite{jk}. This effect may perhaps
also be responsible for the rather large values  of $D$ (exceeding 4)
observed at SPS energies \cite{ce,49}. Data for larger intervals would
be very useful to sort out this question.

(ii) An  argument similar to that presented here can  be  also applied
to fluctuations of the net baryon  number \cite{mu1,mu2}. 
The formulae (\ref{2}) - (\ref{6}) are valid with the substitution of 
the charges by baryon numbers. In particular, for the hadron gas we 
have
\ba
{<\delta B^2>} = \sum_i B_i^2 <n_i> = <N_B>  \l{16}
\ea
whereas for a  parton system one obtains
\ba
{<\delta B^2>} = \frac19 <N_q> .  \l{17}
\ea
Thus   the ratio ${<\delta B^2>}/<N_{ch}> \approx 0.1$ for the QGP while
it is expected to be about three times larger in the coalescence model.
Again, the corrections for resonance decays must be included to obtain
a realistic estimate.

(iii) The constituent quark coalescence model discussed here can also be intepreted as a quark-gluon plasma far from equilibrium. This poses an apparent 
problem, because it has been shown in numerous papers \cite{str} that the equilibrium scenario agrees very well with the data, in particular with those on particle ratios (chemical equilibrium). As I have already mentioned, however,  particle ratios were
analysed in detail by the Budapest group \cite{al,hung} and the coalescence model
was also found to be in excellent agreement with data. I thus feel that it deserves a serious attention. Clearly, however, more work is needed  to clarify fully the situation.

\vspace{0.3cm}
{\bf Acknowledgements}
\vspace{0.3cm}

I would like to thank K.Fialkowski, M.Jezabek, V.Koch and M.Nowak for
discussions. This investigation was supported in part by 
Subsydium of Foundation for Polish Science NP 1/99.

\end{document}